\documentstyle[prl,aps,epsf,multicol]{revtex}
\begin{document}
\draft
\title{Ferromagnetism and superstructure in  Ca$_{1-x}$La$_x$B$_6$}
\author{Victor Barzykin$^{\dagger}$ and Lev P. Gor'kov$^{\dagger,*}$}
\address{$^{\dagger}$National High Magnetic Field Laboratory, 
Florida State University,\\
1800 E. Paul Dirac Dr., Tallahassee, Florida 32310 \\
and \\
$^*$L.D. Landau Institute for Theoretical Physics,
Chernogolovka, 142432, Russia
}
\maketitle
\begin{abstract}
We critically investigate the model of a doped excitonic insulator, which
recently has been invoked to explain some experimental properties of the 
ferromagnetic state in Ca$_{1-x}$La$_x$B$_6$. We demonstrate that the ground state 
of this model is intrinsically unstable towards
the appearance of a superstructure. In addition, the model would lead to a 
phase separation regime and the domain structure which may be prevented 
by the Coulomb forces only. Recent experiments indicate
that a superstructure may indeed show up in this material.
\end{abstract}
\vspace{0.15cm}


\begin{multicols}{2}
\narrowtext
The discovery of weak ferromagnetism in lightly-doped hexaborides
Ca$_{1-x}$La$_x$B$_6$\cite{Young} has renewed theoretical interest
in the so-called ``excitonic'' transition vigorously discussed in the
60-s and 70-s \cite{Arkhipov,KK,KM,French,Kohn}(see \cite{HR} for a review).
Band structure calculations\cite{Massida} indicate that hexaborides, 
DB$_6$, may, in fact, be semimetals owing to
an accidental small band overlap at the three X-points of the Brillouin
Zone, two bands at each X-point having symmetry $X_3$, $X_3'$. It was
suggested\cite{Zhitomirsky} that the use of the model\cite{KK} together
with the energy mechanism for ferromagnetism first suggested in Ref.\cite{VKR,VRT}
provide at least a qualitative explanation for the surprising findings\cite{Young}.
Bearing much of the resemblance with the mathematics of the BCS weak-coupling 
theory of superconductivity, the Keldysh-Kopaev model\cite{KK} is a convenient 
tool which makes possible to treat an excitonic transition in the controllable manner.

Assuming that the basic physics of hexaborides is properly accounted for by this 
oversimplified weak coupling scheme, in what follows we study in some more details
the zero temperature phase diagram as a function of the doping level, and its' stability to
anisotropy features.  The main result is
that the system almost inevitably develops a superstructure on the background of the initially
cubic lattice. Our explanation for the magnetic moment per doped lanthanum ion
to vary and even become very small differs from the one suggested in Ref.\cite{Zhitomirsky}.

Zhitomirsky {\em et al.}\cite{Zhitomirsky} have used the model of Ref.\cite{KK} 
of an excitonic phase transition in a semimetal to explain weak 
ferromagnetism observed in hexaborides. Ferromagnetism is known to appear in this model, and
was investigated  in detail by Volkov {\em et al.} \cite{VKR,VRT}. We first
briefly recall the mechanism of  magnetic moment formation in this model.  
It is well known\cite{KM} that if the screened Coulomb interaction between an electron
and a hole in two bands,
\end{multicols}
\widetext
\begin{equation}
H_{int} = \sum_{\bf k, k', q} \sum_{\alpha \beta} V({\bf q}) 
a_{1 \alpha}^{\dagger}({\bf k}+{\bf q})a_{2 \beta}^{\dagger}({\bf k'}-{\bf q})
a_{2 \beta}({\bf k'}) a_{1 \alpha}({\bf k}),
\end{equation}
\begin{multicols}{2}
\narrowtext
\noindent
is dominant, and the electron and hole Fermi surfaces do coinside (``nesting''),
the excitonic transition has a degeneracy for the onset of CDW (``singlet'') 
and SDW(``triplet'') excitonic condensate. This degeneracy is lifted only by 
additional weak short-range
Coulomb terms, which favor triplet excitonic condensate\cite{HR,Zhitomirsky}, 
or electron-phonon interactions, which favor singlet (CDW) state. This
splitting, however, being usually considered to be weak at
$\delta g \sim g^2$, where $g=V(0) N_0 \ll 1$ is the screened Coulomb coupling constant,
$N_0=m k_F/(2 \pi^2)$ is the density of states per spin for a single band, the temperatures
of the triplet or singlet excitonic transitions are close (on the exponential
scale, $T_c \sim exp[-g^{-1}]$), $T_{s0} \sim T_{t0}$.
Then, as it was first shown in Ref.\cite{VKR,VRT} and applied to the physics of
hexaborides in Ref.\cite{Zhitomirsky}, ferromagnetism can appear as a result of doping due to the
development of a triplet excitonic instability in the presence of a singlet order, or vice versa.
Ferromagnetism is the direct result of the fact that in the presense 
of these two condensates both
crystalline and time-reversal symmetries get broken. Summation of the leading logarithmically 
divergent terms in the presence of one condensate explicitly 
confirms the divergent Curie-Weiss behavior of the spin susceptibility with
some Curie temperature found in Ref.\cite{VRT}.

For the analysis of ferromagnetism in the doped state at low temperatures one may
neglect the small differences in the energy spectrum of the SDW or 
CDW ground state\cite{Zhitomirsky}.
The energetic analysis is straightforward in case when triplet and singlet coupling
constants are equal\cite{Zhitomirsky}, since then equations for different spin 
polarizations are decoupled. Indeed, for a simple model with two
isotropic bands, $m_e = m_h = m$, and 
$\epsilon_e({\bf k}) = {{\bf k}^2 \over  2 m} -  \mu + {E_g \over 2}$,
$\epsilon_h({\bf k}) = {{\bf k}^2 \over  2 m} +  \mu + {E_g \over 2}$,
the zero temperature excitonic gap in the unpolarized state is given by
\begin{equation}
\Delta_{\alpha}^2 = \Delta_0(\Delta_0 - 2 n),
\label{delta}
\end{equation}
where $\Delta_0=2 \epsilon_c exp(-g^{-1})$ is the excitonic gap at zero doping,
$\epsilon_c$ is a cutoff energy around the Fermi surface, $n=N/4 N_0$ is the concentration
of doped carriers in energy units (the level of the chemical potential in 
the bare metallic phase is $\mu=n$). The ground state energy per spin direction 
relative to the stoichiometric normal state is given by:
\begin{equation}
E(n) = N_0 n^2 - {1 \over 2} N_0 (\Delta_0 - 2n)^2.
\label{energy}
\end{equation}
Thus, the excitonic pairing disappears at $n=0.5 \Delta_0$. A quick analysis\cite{Zhitomirsky}
shows that the energy of a polarized excitonic state, $E[n(1+p)] + E[n(1-p)]$ has
a minimum value for $p=1$, i.e. for the complete polarization of added carriers. 
Therefore, below $T_c$, there are two order parameters, 
$\Delta_s$ and $\bbox{\Delta}_t$, and, correspondingly, two different gaps in the spectrum for
different spin polarizations:
\begin{eqnarray}
\Delta_{\downarrow} = \sqrt{\Delta_0(\Delta_0-4n)} \ \ \ 0 < n < \Delta_0/4 \nonumber \\
\Delta_{\uparrow} = \Delta_0  \ \ \ 0 < n < \Delta_0/2.
\end{eqnarray}
Hence at T=0
electrons and holes are paired for only one spin direction when $\Delta_0/4 < n < \Delta_0/2$.
(The system undergoes a 1-st order phase transition into the unpolarized normal metal at 
$n_{cr} = \Delta_0/2$.)
Thus, this mechanism gives a large effective moment equal to 1$\mu_B$ per doped $La$ atom,
and some efforts have been applied in Ref.\cite{Zhitomirsky} to argue that this moment may be
forced to become small if the interaction between the orientation of the magnetic moment 
and the direction ${\bf d}$ of the triplet order parameter (SDW), 
$\bbox{\Delta}_t(p) = ({\bf d}(p) \bbox{\sigma})$, was taken into account. The mechanism
of Ref.\cite{Zhitomirsky} seem to us too artificial, first of all, because the energy
difference between CDW and SDW states was ignored. Meanwhile, 
it is easy to see that an anisotropy of electron and hole pockets would reduce the net 
magnetization.
Indeed, the two opposite spin polarizations preferred by the system are governed by the
position of the chemical potential. First, an anisotropic solution for the order parameter,
$\Delta({\bf p})$, would result in variation of the gap itself along the Fermi surface. Secondly, 
even a small anisotropy (``antinesting'') of the electron and hole spectra hinders the excitonic
gap, even leading to the gapless pockets along the initial Fermi surface \cite{Kopaev,Zittartz}.
Therefore, while starting from the spin-up and spin-down spectra for the isotropic gap
shown in Fig.1a
(1 $\mu_B$ per La), one would end up with doped electrons to spill over between the two energy
branches thus obviously reducing the total magnetization, as it is shown in Fig. 1b.  
\begin{figure}
\epsfxsize=3 in
\epsfbox{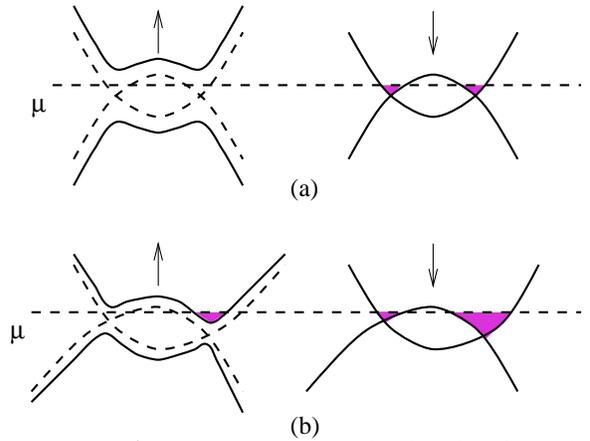}
\caption{a) Energy bands and spin directions in the isotropic excitonic insulator;
b) the value of 
the total magnetic moment depends on the shape of the energy spectrum in momentum space along
the Fermi surface when the gap anisotropy and/or ``antinesting'' are taken into account.}
\label{fig1}
\end{figure}

We will not pursue the detailed calculations for these mechanisms, since the weak coupling  
model\cite{KK} of an excitonic transition
suffers from several obvious defficiencies, each of which results in an instability 
towards a formation of
an inhomogeneous structure. First of all, for the homogeneous excitonic state to exist, the
electron and hole Fermi surfaces should be sufficiently close to nesting. Indeed, a quick
calculation shows that in a model with mass anisotropy, 
\begin{eqnarray}
\epsilon_h({\bf p}) &=& {p_x^2 \over 2m + \delta m} + 
{p_y^2 \over 2m - \delta m} + {p_z^2 \over 2m} + (1/2) E_g \nonumber \\
\epsilon_e({\bf p}) &=& {p_x^2 \over 2m - \delta m} + 
{p_y^2 \over 2m + \delta m} + {p_z^2 \over 2m} + (1/2) E_g,
\end{eqnarray}
a homogeneous excitonic phase disappears at
\begin{equation}
{\delta m \over m} > {2 e \Delta_0 \over |E_g|},
\label{range}
\end{equation}
where $\Delta_0$ is the excitonic gap for the isotropic situation ($\delta m = 0$);
$e =2.71828$. Since $\Delta_0$ is exponentially small in terms of $\epsilon_c \sim E_g$, 
$\Delta_0=2 \epsilon_c exp(-1/g)$, the weak coupling model 
allows only a modest mass anisotropy. 

The range of anisotropy for exciton formation
may be extended beyond Eq.(\ref{range}), however, since at larger anisotropies it
leads to an inhomogeneous state with long-wavelength oscillations of spin and 
(or) electron densities\cite{GM}. 
Similar drawback of the Keldysh-Kopaev model becomes obvious when one considers doping
even in a model with the perfect nesting. Mathematically the problem in the last case 
becomes equivalent to 
the problem of the coexistence of the superconductivity and ferromagnetism\cite{Baltensperger}. 
The exchange field, $I$, in
that model is equivalent to the chemical potential $\mu$ in the excitonic state.
Thus, the gap equation has two solutions:
1) $\Delta = \Delta_0$; 2) $\Delta^2 = 2 I \Delta_0 - \Delta_0^2$, where $\Delta_0$ is
the gap at $I=0$. The energies of the  ground state are easily found:
1) $\Omega - \Omega_0 = -N_0 \Delta_0^2$; 
2) $\Omega-\Omega_0 = -N_0 (4 \Delta_0 I - \Delta_0^2 - 2 I^2)$. 
The branch (2) is not stable for the homogeneous superconductor\cite{GR}.
The physical difference between the problems of the doped excitonic insulator and the 
ferromagnetic superconductor
is that in case of an excitonic insulator the number of added
dopants is fixed, not the chemical potential. Then the ``stable'' solution (1) corresponds to zero
doping, the case when the chemical potential $\mu$ lies below the gap edge $\Delta_0$. 
The dependence of a homogeneous solution for the gap on doping then is given by (2), with $\mu$ being
expressed in terms of added dopants, $n = \sqrt{\mu^2-\Delta^2}$ (where n is in energy units),
reproduces Eq.(\ref{delta}) and the effects considered by
Volkov {\em et al.}\cite{VKR,VRT} and Zhitomirsky {\em et al.}
\cite{Zhitomirsky}. 

The instability of the homogeneous solution at larger $n$ is seen from the form of $T_c(n)$,
explicitly calculated (see Fig. 2) in Ref.\cite{GM}:
\end{multicols}
\widetext
\begin{equation}
T_c = {\gamma \Delta_0 \over \pi} exp\left[\Psi\left({1 \over 2}\right) - 
{1 \over 2} \Psi\left({1 \over 2} + i {n \over 2 \pi T_c}\right) - 
{1 \over 2} \Psi\left({1 \over 2} - i {n \over 2 \pi T_c}\right)\right]. 
\label{TC}
\end{equation}
\begin{multicols}{2}
\narrowtext
\noindent
At high enough densities, $n > \Delta_0/2$, $T_c(n)$ displays a reentrant behavior from the 
excitonic insulator into the metallic state at low temperatures. The behavior indicates
some sort of an instability or a 1-st order phase transition.  
However, similar to the case of non-perfect nesting above, one can again check for an 
instability towards 
a transition into an excitonic state with an incommensurate wave vector ${\bf q}$, searching for 
a maximum $T_c(|{\bf q}|)$. 
\begin{figure}
\epsfxsize=3 in
\epsfbox{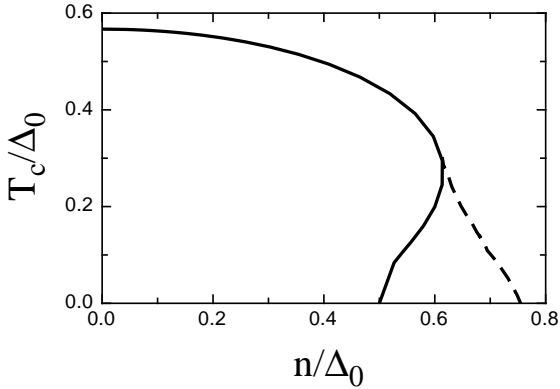}
\caption{Transition temperature into the excitonic state as a function of the concentration 
of dopants, $n=N/(4 N_0)$. The dashed line shows the phase transition into inhomogeneous
excitonic insulator with the wave vector ${\bf q}$.}
\label{fig2}
\end{figure}
The result of our numerical calculation is shown in Fig.2. At low dopings  $T_c$ 
corresponds to the homogeneous
state with ${\bf q} = 0$. The dashed line in Fig. 2 shows the temperature for the onset of
an inhomogeneous phase, $T_c(|{\bf q}|)$, where the value of $|{\bf q}|$ itself depends on $n$.
(At low $T_c$ $|{\bf q}|=2.4 n/v_F$.) The excitonic regime at $T=0$ for such 
inhomogeneous state appears by a first order transition at $n^* = 0.71 \Delta_0$
(see below) and
extends to $n_{c1} \simeq 0.755 \Delta_0$.
(Inhomogeneous state was first discussed by Rice in connection with itinerant
antiferromagnetism in chromium\cite{Rice}.)
 
Assuming again that CDW- and SDW- states have equal energies ($\delta g = 0$),
it is easy to show that $T_c$
in the presence of magnetic field, $B$, initially increases with $B$, as 
follows from $T_c(n,B)$ obtainable by a mere substitution:
\begin{equation}
T_c(n,B) \equiv T_c(n-\mu_B B),
\label{subst}
\end{equation}
from Eq.(\ref{TC}). The $|{\bf q}|$-value in the presence of the magnetic 
field $B$ also follows from the 
same substitution. This result may also be considered as another 
manifestation of the system's tendency towards a
ferromagnetic state. According to Eq.(\ref{subst}), at a higher doping the 
magnetic field may cause reentrance into the excitonic insulator phase.

The vicinity of the upper concentration, $n_{c1}$,
where the non-homogeneous solution first appears
if the concentration is {\em decreased} from the side of the normal metal,
may be studied in the same manner
as for the corresponding superconductivity
problem, producing the well-known
Larkin-Ovchinnikov-Fulde-Ferrel state\cite{LO,FF}. The solution in our
case (we need to find $\mu(n)$) again secures the stripe phase as the most energetically
favorable one.  The density of carriers then oscillates according to:
\begin{equation}
n(x) = n - n [\Delta_b^2/n_{c1}^2] (1 + 1.3 cos(2 q x)),
\label{charge}
\end{equation}
where $q v = 2.4 (n_{c1} + \delta n/15.7)$, $\Delta(x) = 2 \Delta_b cos(q x)$,
and 
\begin{equation}
|\Delta_b|^2 = 0.936 n_{c1} (n_{c1} - n).
\end{equation}
Here $v$ is the Fermi velocity.

An inhomogeneous distribution Eq.(\ref{charge}) of a weak enough charge on the scale of a 
coherence length $\xi_0 > \xi_{TF}$,
where $\xi_{TF}$ is the Thomas-Fermi screening radius should not present a major trouble, since the
lattice can adjust itself to produce a periodic modulation. 
Yet it remains unclear how at low $T$ the system evolves when one proceeds from the 
metallic end by further  decrease of the concentration of dopants.

In any event, if the Coulomb forces are completely neglected, with further
concentration decrease the homogeneous phase is restored through the phase separation
regime. This starts to take place at the point
$n^* = \Delta_0/\sqrt{2}$\cite{GR}.
To find the two phases and how they coexist, in our case turns out 
to be completely equivalent to minimizing the energy of the intermediate state of the 
$I^{st}$-type superconductor in a fixed magnetic field.
The same energy considerations show that the system separates into the mixture of two phases: 
the one with $n_e = n_h$ and the other with $n_e - n_h = n^* = \Delta_0/\sqrt{2}$. The 
relative volume fraction of each phase is given by:
\begin{equation}
V_1/V = 1 - n/n^* \ \ \ V_2/V = n/n^*.
\end{equation}
The domain sizes would be determined by the surface energy.
Without the bulk Coulomb forces such phase separation is energetically even
more favorable than ferromagnetism.
On the other hand, the Coulomb energy would work against the spatial charge 
separation, which, hence, may stabilize a homogeneous
regime in a range of small enough $n$. In view of these uncertainties we
made no attempt to address the other issue, namely, that for a complete description
of the true ground state of DB$_6$ in the frameworks of the model\cite{KK,Massida},
one has to account for the three
X-points in the cubic system related to each other by the symmetry transformations.
This has been done recently in Ref. \cite{ABG} for the problem of {\em multiband
superconductors}, and the results of Ref.\cite{ABG} are immediately mapped onto
the problem of excitonic insulator with the spectrum of Ref.\cite{Massida}.

In conclusion, we have shown that the excitonic model\cite{KK}, 
even though may capture a qualitatively correct physics, necessarily leads to 
appearance of a superstructure as far as the doping dependence is concerned.
We are aware of the only experimental observation yet of such a 
superstructure\cite{Experiment} which, at least, does not contradict our expectations
above.
We emphasize again that our explanation for possible small magnetic moment differs
from the one given
in Ref.\cite{Zhitomirsky}. The magnetic moment can be as big as $\sim 1 \mu_B$ 
per doped $La$ in the model\cite{VKR}. Effects of anisotropy and changes in the 
energy spectrum through doping would change this significantly, as mentioned above.
Recent experiments\cite{Zach} on 
$BaB_6$ doped  with $La$ have produced moments about $0.4 \mu_B$
at $x=0.05$.

After this work was performed, the authors have discovered the preprint Ref.\cite{BV}  
by Balents and Varma. We shall not discuss their results 
regarding the three different $X$-points. We note that our results 
on the doping dependence agree for the most part. We do not agree, however, with the
claim that the results of \cite{VKR,VRT,Zhitomirsky} contain significant errors,
since the macroscopic phase separation is to be made difficult by the Coulomb terms.

The authors thank Z. Fisk for everyday discussions on both the 
experimental results and related physics. We are grateful to 
R. Caciuffo for communicating the results of Ref.\cite{Experiment} prior
to publication. L.P.G. also acknowledges with gratitude stimulating discussions 
with M. Chernikov, A. Bianchi, S. Oseroff, H.-R. Ott, and C. M. Varma.
This work was supported by the National High Magnetic
Field Laboratory through NSF cooperative agreement 
No. DMR-9527035 and the State of Florida.

\end{multicols}
\end{document}